\documentclass[sigconf]{acmart}

\usepackage{epsfig}
\usepackage{graphicx,subfigure}
\usepackage{verbatim}
\usepackage{amsmath}
\usepackage{amssymb}
\usepackage{float}
\usepackage{multicol}
\usepackage{times}
\usepackage{multirow}
\usepackage{rotating}
\usepackage{bm}
\usepackage{diagbox}
\usepackage{booktabs}
\usepackage{amsmath,epsfig}
\usepackage{amssymb}
\usepackage{epstopdf}
\usepackage{graphicx,subfigure}
\usepackage{enumerate}
\usepackage{amsmath}
\usepackage{amscd}
\usepackage{rotating}
\usepackage{multirow}
\usepackage{url}
\usepackage{extarrows}
\usepackage{booktabs}
\usepackage{times}
\usepackage{balance}
\usepackage{color}
\usepackage{colortbl}
\usepackage{algorithmic}
\usepackage{algorithm}

\usepackage{booktabs} 

\setcopyright{rightsretained}

\acmDOI{10.475/123_4}

\acmISBN{123-4567-24-567/08/06}

\acmConference[KDD'18]{ACM SIGKDD}{Aug 2018}{London, United Kingdom}
\acmYear{2018}
\copyrightyear{2018}

\acmPrice{15.00}

\begin{document}
\title{Cross-domain Novelty Seeking Trait Mining for Sequential Recommendation}


\author{Fuzhen Zhuang}
\affiliation{%
  \institution{IIP, ICT, CAS, Beijing, China}
}
\email{zhuangfuzhen@ict.ac.cn}

\author{Yingmin Zhou}
\affiliation{%
  \institution{IIP, ICT, CAS, Beijing, China}
}
\email{zhouyingmin@ict.ac.cn}

\author{Fuzheng Zhang}
\affiliation{%
  \institution{Microsoft Research Asia}
}
\email{fuzzhang@microsoft.com}

\author{Xiang Ao}
\affiliation{%
  \institution{IIP, ICT, CAS, Beijing, China}
}
\email{aoxiang@ict.ac.cn}

\author{Xing Xie}
\affiliation{%
  \institution{Microsoft Research Asia}
}
\email{xingx@microsoft.com}

\author{Qing He}
\affiliation{%
  \institution{IIP, ICT, CAS, Beijing, China}
}
\email{heqing@ict.ac.cn}

\renewcommand{\shortauthors}{F.Z. Zhuang et al.}

\begin{abstract}
Transfer learning has attracted a large amount of interest and research in last decades, and some efforts have been made to build more precise recommendation systems. Most previous transfer recommendation systems assume that the target domain shares the same/similar rating patterns with the auxiliary source domain, which is used to improve the recommendation performance. However, to the best of our knowledge, almost these works do not consider the characteristics of sequential data. In this paper, we study the new cross-domain recommendation scenario for mining novelty-seeking trait. Recent studies in psychology suggest that novelty-seeking trait is highly related to consumer behavior, which has a profound business impact on online recommendation. Previous work performing on only one single target domain may not fully characterize users' novelty-seeking trait well due to the data scarcity and sparsity, leading to the poor recommendation performance. Along this line, we proposed a new cross-domain novelty-seeking trait mining model (CDNST for short) to improve the sequential recommendation performance by transferring the knowledge from auxiliary source domain. We conduct systematic experiments on three domain data sets crawled from Douban (www.douban.com) to demonstrate the effectiveness of the proposed model. Moreover, we analyze how the temporal property of sequential data affects the performance of CDNST, and conduct simulation experiments to validate our analysis.
\end{abstract}

%
%
\vspace{-0.5cm}



\vspace{-0.4cm}

\maketitle

\vspace{-0.3cm}
\section{Introduction} \label{sec:introduction}

Personalized recommendation plays a very important role in the rapid development of E-commerce. To make more precise recommendations for personal needs, we should understand users' preference propensity or profiles according to their historical behaviors. For example, on the well known E-commerce website Amazon, we may make recommendations to one user if he shares the similar consuming behaviors with other ones, or according to his historical consuming behaviors. Therefore, recommendation system has attracted vast amount of interest and research in recent years to handle the information overload problem and  make predictions~\cite{Bobadilla2013109,SuSurvey2009}.

Unlike most of previous works, there has been some effort devoted to modeling an individual's propensity from psychological perspective for recommendation systems in recent years~\cite{ZhangYLX2014,ZhangZYX2015}. Novelty seeking is a personal trait described as the search for unfamiliar experiences and feelings that are ``varied, novel, complex, and intense'', and measured by the readiness to take ``physical, social, legal, and financial'' risks for the sake of such experiences. Novelty seeking, as well as harm avoidance and reward dependence, has been regarded as the basic requirement for human activities~\cite{Cloninger94}.  Behaviors of users are also relatively consistent in similar situations~\cite{furr2004situational}.  In consumer behavior and recommender system research, understanding this personality trait is particularly crucial since consumers' attributes are strong indicators of their purchasing behaviors~\cite{stern1962significance}. Hence, if you know more about whether your consumer loves trying new things, you can recommend your product more reasonably according to consumer's taste and reach your targets faster and more effectively. To that end, Zhang et al.~\cite{ZhangYLX2014} proposed a computational framework named Novel Seeking Model (NSM) to explore the novelty-seeking trait implied by observable sequential activities. Experimental results showed that NSM can uncover the correlation of novelty-seeking trait at different levels, and improve the recommendation performance. Following this line, Zhang et al.~\cite{ZhangZYX2015} also proposed a novelty-seeking based dining recommendation system for effective dining recommendation.

Users are always active in many E-commerce websites, and have large number of sequential behavioral data in different domains. As we all know, a user's behaviors in different areas have consistency. For example, if a user watches movies focused on his favourite actors, this phenomenon shows that the user is lower novelty-seeking propensity, so he will listen some particular genre of music. The modeling of novelty-seeking trait in one single domain may not completely characterize each individual's profiles, while the sequential behavioral data of one user from different domains may help to exploit the novelty-seeking trait. For example, on the well known Chinese social media platform Douban\footnote{https://www.douban.com/}, users usually read books, listen to music, watch movies, and then express their propensity comments. Observing these three domains of sequential behavioral data, we find that users listened some music and then after a period of time they would watch some related movie, e.g., the music is the theme music of the movie; users sometimes watch some movies after they read some related books, from which the movies are derived. Based on these observations, whether the sequential behavioral data of domains of Book and Music can help to model the novelty-seeking trait in the domain of Movie? This issue is crucial to cross-domain recommendation, especially in the situation where one domain suffers from the cold-start problem. On the other hand, transfer learning aims to transfer the knowledge from related auxiliary source domain to target domain. Along this line, we propose a new cross-domain novelty-seeking trait mining model, termed as CDNST, in which the parameters characterizing the novelty-seeking trait are shared across different domains to achieve significant improvement for recommendations. We crawled three domains of data from Douban website, i.e., Book, Music and Movie, and conduct extensive experiments to validate the effectiveness of the proposed model. The experiments also indicate that performance of CDNST is sensitive to the sequential property of related domain data, which inspires us to study new cross-domain method in the future.

Our contribution can be summarized as follows,
\begin{itemize}
  \item We propose a new cross-domain novelty-seeking algorithm for better modeling an individual's propensity from psychological
perspective for recommendation, in which the novelty-seeking level of each individual is shared for knowledge transfer across different domains.
  \item We crawl three domains of data sets from the well-known Chinese social-media platform Douban and construct 14 transfer recommendation problems, to demonstrate the effectiveness of the proposed model CDNST.
  \item We are the first to analyze how the temporal property of sequential data affects the transfer learning model, and conduct simulation experiments to validate our analysis. Moreover, we define an effective relatedness measure to decide what kinds of transfer learning problems are suitable to our model.
\end{itemize}

The remainder of this paper is organized as follows. Section~\ref{sec:relatedwork} briefly introduces the related work. Section~\ref{sec:model} details the problem formulation and solution derivation of CDNST. The effectiveness and analysis experiments are shown in Section~\ref{sec:experiments}. Finally, Section~\ref{sec:conclusions} concludes this paper.

\vspace{-0.3cm}
\section{Related Work} \label{sec:relatedwork}
In this section, we briefly introduce the most related work on novelty-seeking research and transfer recommendation systems.
\vspace{-0.3cm}
\subsection{Novelty Seeking Research}
Novelty seeking is a personality trait expressed in the generalized tendency to seek varied, novel, complex, and intense sensations and experiences and the willingness to take risks for the sake of such experiences~\cite{zuckerman1979sensation}. Consumer behavior and health science focused on novelty seeking a long time~\cite{ebstein1996dopamine}. It is construed as sensation seeking or neophilia. The notion of novelty seeking was proposed by Acker and McReynolds~\cite{acker1967need}. And then it was studied by McClenland~\cite{mcclelland1955studies}, Fiske and Maddi~\cite{fiske1961functions} and Rogers\cite{rogers2010diffusion}. Rajus~\cite{raju1980optimum} studied personality traits, demographic variables, and exploratory behavior in the consumer context. Baumgartner~\cite{baumgartner1996exploratory} proposed a two-factor conceptualization of exploratory consumer buying behavior. Zhang~\cite{ZhangYLX2014} presented a computational framework for exploiting the novelty-seeking trait implied by the observable activities. Our work is inspired by \cite{ZhangYLX2014}, and focused on transfer learning model for better recommendation.
\vspace{-0.3cm}
\subsection{Transfer Recommendation Systems}
In order to integrate more information from different domains for better recommendation,  cross-domain recommendation considers to combine data sources from different domains with the original target data~\cite{LiYX09Transfer,TCKR12}.  The basic idea of existing methods utilize the common latent structure shared across domains as the bridge for knowledge transfer. Recently, considering the number of overlapped users is often small,  Jiang et al.~\cite{JCYXY16}  proposed a novel semi-supervised transfer learning method to address the problem of cross-platform behavior prediction. Wei et al.~\cite{wei2017beyond} proposed a Heterogeneous Information Ensemble framework to predict users' personality traits by integrating heterogeneous information including self-language usage, avatar, emotion, and responsive patterns. Lian et al.~\cite{lian2017cccfnet} proposed CCCFNet which combine collaborative filtering and content-based filtering for cross-domain Recommendation. Wang et al.~\cite{wang2017dynamic} proposed a model across multiple deep neural nets to catch representation learning of each article and capture the change.

Although these cross-domain recommendation methods have achieved successes in many applications, these methods are usually designed for statical rating data. This paper focuses on the transfer recommendation system for sequential behavior data from psychological perspective. 
\vspace{-0.3cm}
\section{Model and Solution} \label{sec:model}

In this section, we present a cross-domain framework to explore the novelty-seeking trait embodied in an individual's behavioral data in a target domain by transferring knowledge from related auxiliary source domain data. First, we would like to clarify some of the
notions commonly used in this paper, and then propose the cross-domain novelty-seeking trait mining model (CDNST). Finally, the solution of CDNST is inferred.
\vspace{-0.3cm}
\subsection{Preliminaries}
\begin{figure}[!th]\vspace{-0.4cm}
\centering
\includegraphics[scale=0.45]{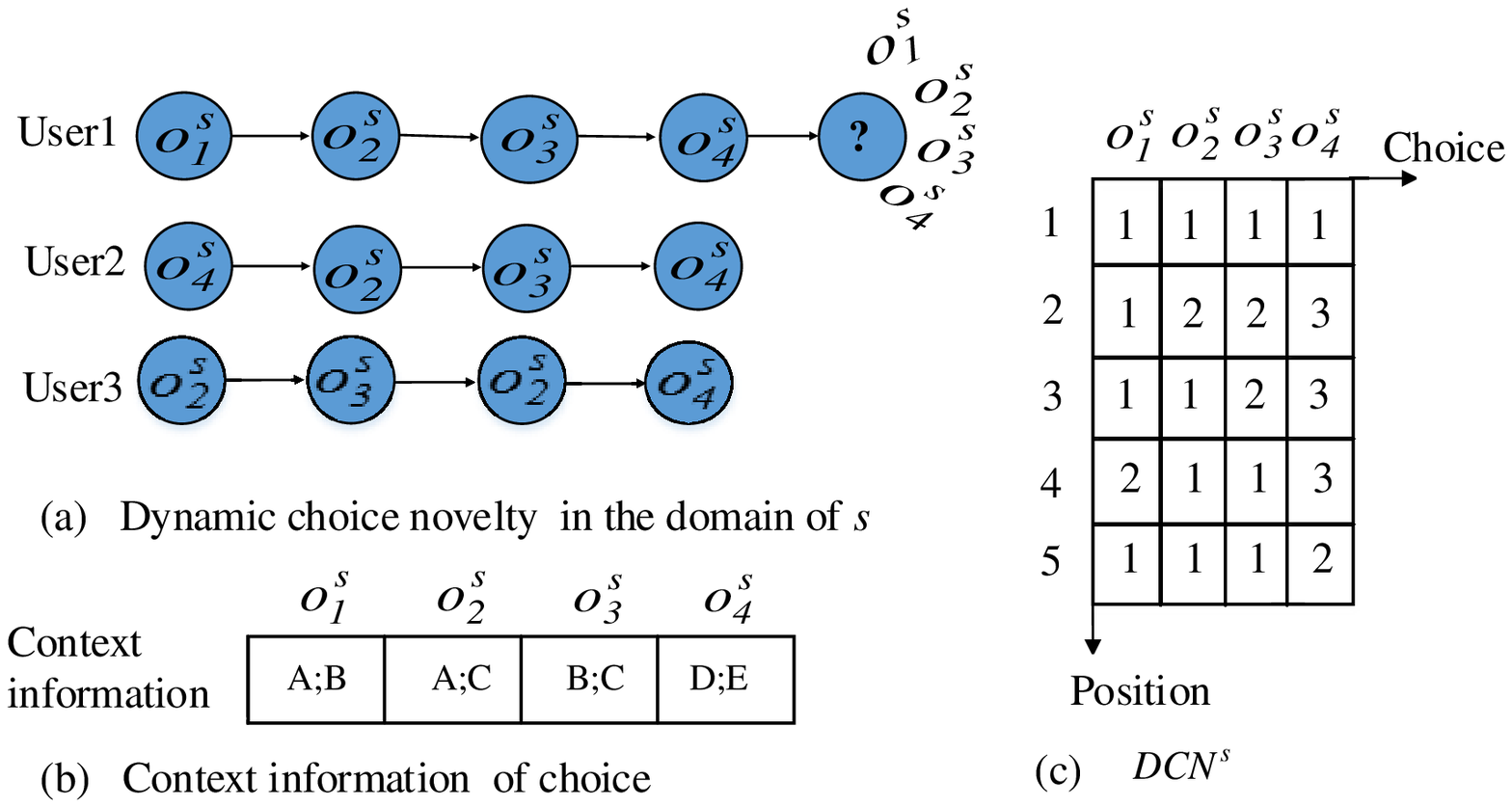} \vspace{-0.4cm}
\caption{Dynamic choice novelty with regards to individual in the domain of $s$. }
\label{fig} \vspace{-.3cm}
\end{figure}

$\mathbf{Action}$ and $\mathbf{Choice}$: Denote $x^{s}$ and $x^{t}$ the specific observed behavior taken by an individual in the source domain $s$ and target domain $t$, respectively. Meanwhile, $x^{s}$ and $x^{t}$ are separately selected from their optional choices $\mathbb{O}^{s}$ and $\mathbb{O}^{t}$, i.e., $x^s\in \mathbb{O}^{s} = { \{o_1^s,\cdots,o_{M_s}^s\} }$ and $x^t \in { \mathbb{O}^{t} = \{o_1^t,\cdots,o_{M_t}^t\}}$, where
$M_s$ and $M_t$ are numbers of choices in the domains of $s$ and $t$, respectively.
The granularity of choices can vary according to the different data format and applications. For example, an \emph{action} on Amazon refers to the purchase of an item, where the \emph{choices} are all available items. For Douban, in particular, an \emph{action} could refer to comment for a specific artwork, which is considered as one of the \emph{choices}.
Besides, every choice candidate in both domain $s$ and $t$ has its context information, which involves categories, tags, keyword, and etc. For example, the information of players, e.g., ``Will Smith'', ``Tony Stark'', could be found in the keywords of a movie~(choice) in Douban movie channel; the category of music~(choice), e.g., ``folk'', ``R\&B'', are presented as tags in Douban music channel. The action sequence ${\textbf{x}^{\cdot}}$ $=$ $(x_1^{\cdot}, x_2^{\cdot}, \cdots, x_{N}^{\cdot})$ of an individual refers to the actions taken in chronological order in a specific domain, where $N$ is the number of actions. We show a general example of action and choice in Fig.~\ref{fig}~(a) and (b). In more detail, Fig.~\ref{fig}~(a) demonstrates three users' action sequences, and Fig.~\ref{fig}~(b) exhibits four choice candidates in the running example, in which ``A'', ``B'', ``C'',``D'' and ``E'' denote the context information of these choices.

\begin{table}[htbp] \vspace{-0.3cm}
 \centering
 \caption{Summary table of symbols}\label{tb:notation} \vspace{-0.3cm}
	\begin{tabular}{@{}lll@{}}
		\hline
		Notations & Denotations  \\
		\midrule
		$K$ & number of optional values for novelty \\ &-seeking level  \\
		$M_s$ & number of optional choices for an action \\& in the domain of $s$ \\
		$ M_t$ & number of optional choices for an action \\& in the domain of $t$ \\
		$N_s$ & length of  actions in the domain of $s$ \\
		$N_t$ & length of  actions in the domain of $t$ \\
		$\mathbf{x^s}$  = ${\left(x_1^s,x_2^s,...,x_{N_s}^s\right)}$& a vector indicates action \\& sequence in the domain of $s$ \\
		$\mathbf{x^t}$  = ${\left(x_1^t,x_2^t,...,x_{N_t}^t\right)}$& a vector indicates action\\& sequence in the domain of $t$ \\
		$\mathbf{z^s}$  = ${\left(z_1^s,z_2^s,...,z_{N_s}^s\right)}$& a vector indicates novelty-seeking level \\& sequence in the domain of $s$ \\
		$\mathbf{z^t}$  = ${\left(z_1^t,z_2^t,...,z_{N_t}^t\right)}$& a vector indicates novelty-seeking level \\& sequence in the domain of $t$ \\
		$\mathbf{\theta}$  = $\left\{\theta_1,\theta_2,...,\theta_K\right\}$& novelty-seeking level distribution \\
		$\mathbf{\phi^s}$  = $\left\{\phi_1^s,\phi_2^s,...,\phi_{M_s}^s\right\}$& choice utility distribution  in the\\&  domain of $s$ \\
		$\mathbf{\phi^t}$  = $\left\{\phi_1^t,\phi_2^t,...,\phi_{M_t}^t\right\}$& choice utility distribution in the\\&  domain of $t$ \\
		$DCN^s_{N_s \times M_s}$ & dynamic choice novelty matrix in the \\& domain of $s$ \\
		$DCN^t_{N_t \times M_t}$ & dynamic choice novelty matrix in the \\&  domain of $t$ \\
		$\alpha^s,\alpha^t,\beta$ & hyperparameters relate to $\phi^s$ ,$\phi^t$ \\& and $\theta$ separately.\\
		\bottomrule
	\end{tabular} \vspace{-0.4cm}
\end{table}

$\mathbf{Dynamic}$ $\mathbf{Choice}$ $\mathbf{Novelty (DCN)}$: Given an arbitrary domain, $\mathit{DCN}$~\cite{ZhangYLX2014} is a $N$ $\times$ $M$ matrix, where $N$ is the length of action sequences and $M$ denotes the number of choices in such domain. Every element in $\mathit{DCN}$ is an integer in $[1, M]$.
The $\mathit{DCN}$ is used to present partial orders among $M$ choices at each position. In more detail,  the $i$-th row in $\mathit{DCN}$ measures the partial orders among $M$ choices at $i$-th position. For instance, the \textsf{User1} in Fig.~\ref{fig}~(a) has four choices at the $5$-th position in his action sequence, and its corresponding row in $\mathit{DCN^s}$, i.e., the $5$-th row of $\mathit{DCN^s}$ as shown in Fig.~\ref{fig}~(c) refers to the current novelty of such four choice candidates which are related to two factors: (1) popularity of the choice and (2) popularity of the choice transition, given historical observations. The more popular the two factors, the lower ranking the choice. The $\mathit{DCN}$ for a given source domain $s$ for example can be computed according to the following principle
\begin{equation}
DCN^s  \propto  \frac{1}{\left( \# x^s_{i}+1 \right)\cdot \left(T_{x^s_{i-1}x^s_{i}}+1\right)},
\end{equation}
where $\#x_{i}^s$ refers to the frequency of terms in context information~(keyword for specific) in $x_{i}^s$ before the $i$-th position in this individual's sequence. It measures the popularity at that moment in view of this individual. $T_{x_{{i-1}}^sx_{i}^s}$ refers to the transition probability of keywords in $x_{i-1}^s \rightarrow x_i^s$ before $i$-th position in this individual's sequence, which measures the context information transition popularity at the moment in view of this individual. The notation for the target domain $DCN^t$ can be obtained by the similar way.

For example, regarding \textsf{User1} in Fig.~\ref{fig}~(a), the novelty order of choices at $5$-th position is o$_4^s$ \textgreater o$_3^s$ = o$_2^s$ = o$_1^s$ since the frequency of terms in o$_4^s$ is minimal and transfer of o$_3^s$ , o$_2^s$ and o$_1^s$ considering this individual's historical behavior. This is partial order according to Equation~(1), and $DCN_{i, j}^{s}$ denotes the novelty-seeking of $s_{j}^{s}$ taken by a user at position of $i$, thus ${DCN^s_{5, 4}}=2, {DCN^s_{5, 1}}=1, {DCN^s_{5, 2}}=1, {DCN^s_{5, 3}}=1$ (Note that 1 indicates the lowest ranking and vice versa).

$\mathbf{Novelty}$-$\mathbf{Seeking}$ $\mathbf {Level}$: The novelty-seeking level $z$ $\in$ $[1, K]$ is a positive integer, where a larger value indicates a higher novelty-seeking propensity and vice versa in a given domain. In the action sequence of an individual in the domain, each position relates to a specific novelty level, e.g., if \textsf{User1} choose $o_{4}^s$ at the last position in domain $s$, it is more likely he has a high novelty-seeking propensity at that moment and want to explore something new in such domain. Otherwise, $o_{1}^{s}$, $o_{2}^{s}$ and $o_{3}^{s}$ might be his choice.
We argue that the individual's novelty in different domains sometimes have similar traits, we thus can transfer such knowledge between multiple domains.

$\mathbf{Novelty}$-$\mathbf{Seeking}$ $\mathbf{Trait(NST)}$: Novelty-seeking trait is an real number ranging from 1 to $K$, which refers to the mean of a multinomial distribution $\mathbf{\theta}$ $=\{\theta_1,\theta_2,....,\theta_K\}$, where $\theta_K$ refers to the probability of having novelty-seeking level of $K$. 
As~\cite{ZhangYLX2014} introduced, the larger the NST, the greater the novelty-seeking propensity the individual possesses and vice versa.

\vspace{-0.3cm}
\subsection{The Proposed Model}

In the following, we detail the proposed model, which is inspired by NSM proposed by Zhang et al.~\cite{ZhangYLX2014}. In their work, a graphical model expressing how to generate observable actions in one specific domain was proposed. In this paper, however, the proposed CDNST attempts to transfer the novelty seeking traits
learnt from auxiliary source domain for improving the accuracy of recommendation of target domain. Our notation and terminology closely follows standards in~\cite{ZhangYLX2014} and deviates only when necessary.

\begin{figure}[!th]
	\begin{center}
       \includegraphics[scale=0.45]{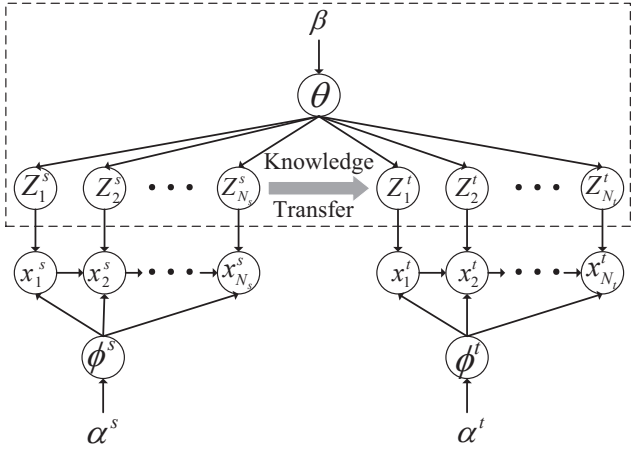}
	\end{center}
	\vspace{-0.5cm}
	\caption{A graphical representation of our general novelty seeking model.}\label{fig:graphicalmodel} \vspace{-0.3cm}
\end{figure}

The notations as well as denotations we use in this model are summarised in Table~\ref{tb:notation}. In CDNST, we extend the framework of NST to multiple domains and give its graphical model as Fig.~\ref{fig:graphicalmodel}. As shown in Fig.~\ref{fig:graphicalmodel}, $z_i^s$ is the latent variable that represents the novelty-seeking level at the position of $i$-th in the source domain $s$. Similarly,  $z_i^t$ denotes the latent variable about the novelty-seeking level at the position of $i$-th in the target domain $t$. Both of them are sampled from a \emph{shared} multinomial novelty-seeking distribution $\mathbf{\theta}$ in Fig.~\ref{fig:graphicalmodel}.
In addition, we use latent variables $\phi^s=\{\phi_1^s,\phi_2^s,...,\phi_{M_s}^s\}$ and $\phi^t=\{\phi_1^t,\phi_2^t,...,\phi_{M_t}^t\}$ to represent the utility of each choice in the domain of $s$ and $t$, respectively. They can be interpreted as this individual's preference for each choice in the corresponding domain. Furthermore, $\alpha^s$, $\alpha^t$ and $\beta$ are the relevant hyper-parameters to ${\phi^s}$, $\phi^{t}$ and ${\theta}$, respectively. The observed actions for the $i$-th position in the domain of $s$ and $t$ are denoted by $x_i^s$ and $x_i^t$ in the figure respectively. The value of $x_i^s$ relies on the novelty-seeking level at $i$-th position, namely $z_i^s$, the choice utility distribution $\phi^s$, and the previous chosen action. The generation process of $x_i^t$ is similar to $x_i^s$ but relies on the corresponding variables in the target domain.

The first-order dependency of the action sequence is still carried out for the different domains in CDNST for simplicity and feasibility. Hence given a dynamic choice novelty matrix $DCN^s$~($DCN^t$) precomputed according to the individual's behavior, and incorporating both the utility and the novelty-seeking factors, the conditional probability is given as:
\begin{eqnarray}
P\left( x_i^s \mid x_{i-1}^s,z_i^s,\phi_s\right)={\frac{\phi_{x_i^s}^s\cdot f\left(z_i^s,DCN_{i, x_{i}^{s}}^s\right)}{\sum_{x_i^s} \left(\phi_{x_i^s}^s\cdot f\left(z_i^s,DCN_{i, x_{i}^s}^s\right)\right)}},
\end{eqnarray}
\begin{eqnarray}
P\left( x_i^t \mid x_{i-1}^t,z_i^t,\phi_t\right)={\frac{\phi_{x_i^t}^t\cdot f\left(z_i^t,DCN_{i, x_{i}^{t}}^t\right)}{\sum_{x_i^t} \left(\phi_{x_i^t}^s\cdot f\left(z_i^t,DCN_{i, x_{i}^t}^t\right)\right)}},
\end{eqnarray}
where the first-order dependency between $x_{i-1}^s$ and $x_{i}^s$  and that between $x_{i-1}^t$ and $x_{i}^t$ are embodied when we compute ${DCN_{i, x_{i}^s}^s}$ and ${DCN_{i, x_{i}^t}^t}$.

The assumption of consistency between the novelty of a choice with the novelty-seeking level at a given position derives that the individual will accept the choice with a higher probability. For instance, if an individual is at the higher novelty-seeking of $K$ at that moment, we expect he/she is more likely to accept a choice with the largest novelty in the partial order. Otherwise, he is likely to accept a choice with little novelty in the partial order. As a result, we give the action function adopted for both source and target domain in CDNST as follows:
\begin{eqnarray}
f\left(z_i^s,DCN_{i, x_{i}^s}^s\right){=\exp\left(-\left(z_i^s-\frac{DCN_{i, x_{i}^s}^s}{\max\left(DCN_i^s\right)}\cdot K\right)^2\right)},\\
f\left(z_i^t,DCN_{i, x_{i}^t}^t\right){=\exp\left(-\left(z_i^t-\frac{DCN_{i, x_{i}^t}^t}{\max\left(DCN_i^t\right)}\cdot K\right)^2\right)},
\end{eqnarray}
where max $\left(\mathit{DCN}_{i}^s \right)$  indicates the maximum value in the $i$-th row of matrix $\mathit{DCN}^s$ , max$\left( DCN_{i}^t \right)$ indicates the maximum value in the $i$-th row of matrix $DCN^t$.

The generative process of CDNST is summarized as Algorithm~\ref{alg:partitionframework}.
\vspace{-0.3cm}
\begin{algorithm}[!th]
\caption{Generative process of CDNST }
\label{alg:partitionframework}
\begin{itemize}
  \item[(1)] Draw novelty-seeking level distribution ${\theta \sim Dirichlet \left( \beta \right)}$;
  \item[(2)] Draw choice utility distribution ${\phi^s \sim Dirichlet \left( \alpha^s \right)}$ in the domain $s$;
  \item[(3)] For the $i$-th position in the sequence
        \begin{itemize}
             \item[(a)] Draw novelty-seeking level $z_i^s$ $\sim$ $\theta$;
             \item[(b)] Draw item  ${x_i^s \sim P\left(x_i^s \mid x_{i-1}^s,\phi^s,z_i^s\right)}$;
            \end{itemize}
  \item[(4)] Draw choice utility distribution${\phi^t \sim Dirichlet \left( \alpha^t \right)}$ in the domain $t$;
  \item[(5)] For the $i$-th position in the sequence
          \begin{itemize}
             \item[(a)] Draw novelty-seeking level $z_i^t$ $\sim$ $\theta$;
             \item[(b)] Draw item  ${x_i^t\sim P\left(x_i^t \mid x_{i-1}^t,\phi^t,z_i^t\right)}$;
            \end{itemize}
\end{itemize}
\end{algorithm}\vspace{-0.5cm}

\subsection{Model Inference}
Following NSM, pointwise Gibbs sampling is applied by repeatedly drawing novelty-seeking level $z^s$ and $z^t$ novelty-seeking level distribution $\mathbf{\theta}$, and choice utility distribution $\phi^s$ and $\phi^t$. The sampling process is summarised as follows:
\begin{itemize}
\item[(1)] Randomly draw $z^s$ from
\begin{eqnarray}
\begin{split}
   P\left(z_i^s\mid \mathbf{z^s_{-i}},x^s,\phi^s,\theta\right)\propto {P\left(\mathbf{z^s},\mathbf{x^s} \mid \phi^s,\mathbf{\theta}\right)}\\
  \propto {\theta_{z_i^s}\cdot \phi_{x_{i}^s}^s \cdot f\left(z_i^s,DCN_{i, x_{i}^s}^s\right)}.
\end{split}
\end{eqnarray}
\item[(2)] Randomly draw $\mathbf{\theta}$ from
\begin{eqnarray}
 \mathbf{\theta}{ \sim Dirichlet\left(\mathbf{\theta} \mid \mathbf{\beta'}\right)},
\end{eqnarray}
where $\mathbf{\beta'}$ is a vector that increases the position $k$ by $n_k$ for $\beta$, $n_k$ is the number of novelty-seeking level with value $k$ in the current state of the sampler.

\item[(3)] Randomly draw $\phi^s$ from
\begin{eqnarray}
\begin{split}
P\left(\phi^s\mid \mathbf{z^s},\mathbf{x_s},\mathbf{\theta} \right)\propto{P\left(\mathbf{x^s}\mid \phi^s,\mathbf{z^s}\right)\cdot P\left(\phi^s\mid \alpha^s\right)}\\
                                        \propto{\frac{\prod_{i=2}^{N_s}\left(\phi_{x_i^s}^s \cdot f\left(z_i^s,DCN_{i, x_{i}^s}^s\right)\right)}{\prod_{i=2}^{N_s}\sum_{x_i^s}\left(\phi_{x_i^s}^s \cdot f\left(z_i^s,DCN_{i, x_{i}^s}^s\right)\right)}}.
\end{split}
\end{eqnarray}
\item[(4)]  Randomly draw $z^t$ from
\begin{eqnarray}
\begin{split}
P\left(z_i^t\mid \mathbf{z^t_{-i}},x^t,\mathbf{\phi^t},\mathbf{\theta}\right)\propto{P\left(\mathbf{z^t},\mathbf{x^t} \mid \phi^t,\mathbf{\theta}\right)}\\
\propto{\theta_{z_i^t}\cdot \phi_{x_{i}^t}^t \cdot f\left(z_i^t,DCN_{i, x_{i}^t}^t\right)}.
\end{split}
\end{eqnarray}
\item[(5)] Random draw $\mathbf{\theta}$ from
\begin{eqnarray}
\mathbf{\theta}\sim {Dirichlet\left(\mathbf{\theta} \mid \mathbf{\beta'}\right)},
\end{eqnarray}
where $\mathbf{\beta'}$ is a vector that increases the position $k$ by $n_k$ for $\beta$, $n_k$ is the number of novelty-seeking level with value $k$ in the current state of the sampler.

\item[(6)] Randomly draw $\phi^t$ from
\begin{eqnarray}
\begin{split}
P(\phi^t\mid \mathbf{z^t},\mathbf{x_t},\mathbf{\theta})\propto{P\left(\mathbf{x^t}\mid \phi^t,\mathbf{z^t}\right)\cdot P\left(\phi^t\mid \alpha^t\right)}\\
                                         \propto{\frac{\prod_{i=2}^{N_t}\left(\phi_{x_i^t}^t \cdot f\left(z_i^t,DCN_{i, x_{i}^t}^t\right)\right)}{\prod_{i=2}^{N_t}\sum_{x_i^t}\left(\phi_{x_i^t}^t \cdot f\left(z_i^t,DCN_{i, x_{i}^t}^t\right)\right)}}.
\end{split}
\end{eqnarray}
\end{itemize} 
\vspace{-0.3cm}
\section{Experiments} \label{sec:experiments}

In this section, we first conduct extensive experiments to demonstrate the effectiveness of the proposed model CDNST, and then analyze how the temporal property of sequential data affects the performance of CDNST. Finally, we design some simulation experiments to validate our analysis and define an effective relatedness measure to judge what kinds of transfer learning problems are suitable to our model.
\begin{figure}[!th] \centering \vspace{-0.4cm}
\subfigure[An user's watching list of movies.] { \label{figexample:a}
\includegraphics[width=0.47\columnwidth]{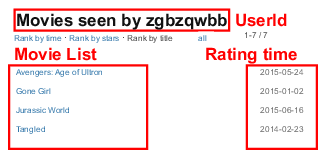}
}
\subfigure[An example of movie's information.] { \label{figexample:1}
\includegraphics[width=0.47\columnwidth]{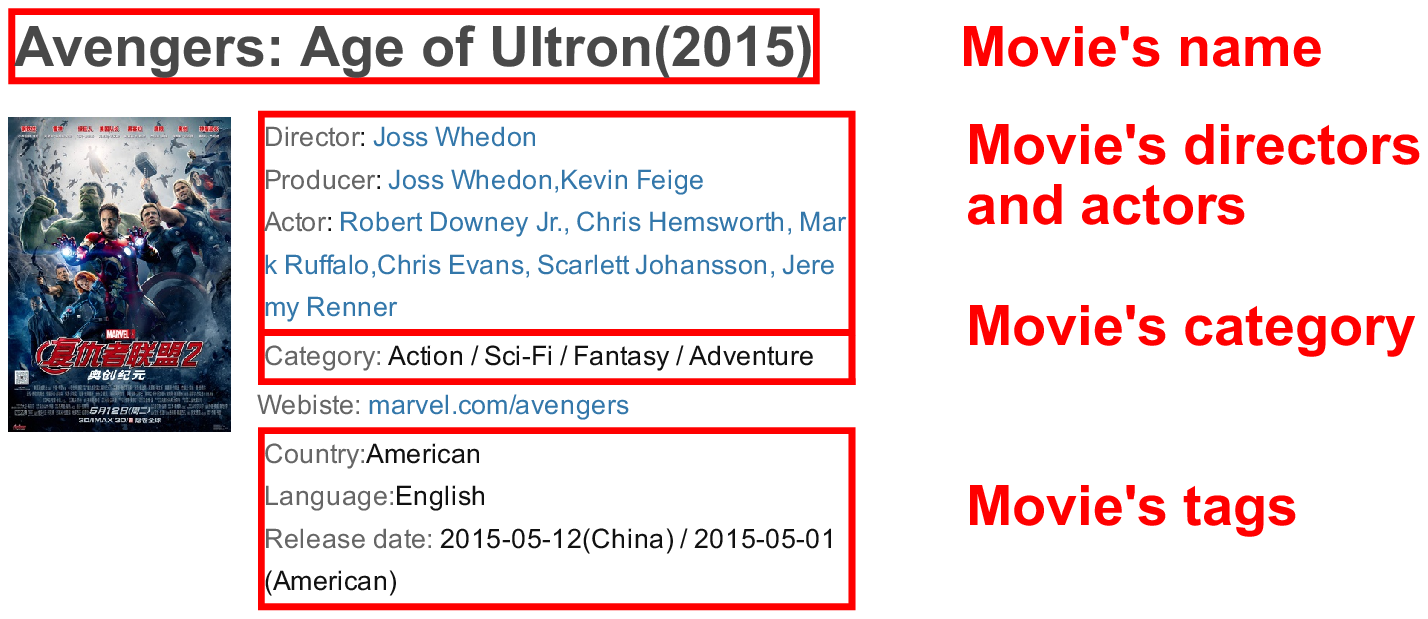}
}
\subfigure[An example of music's information.] { \label{figexample:2}
\includegraphics[width=0.47\columnwidth]{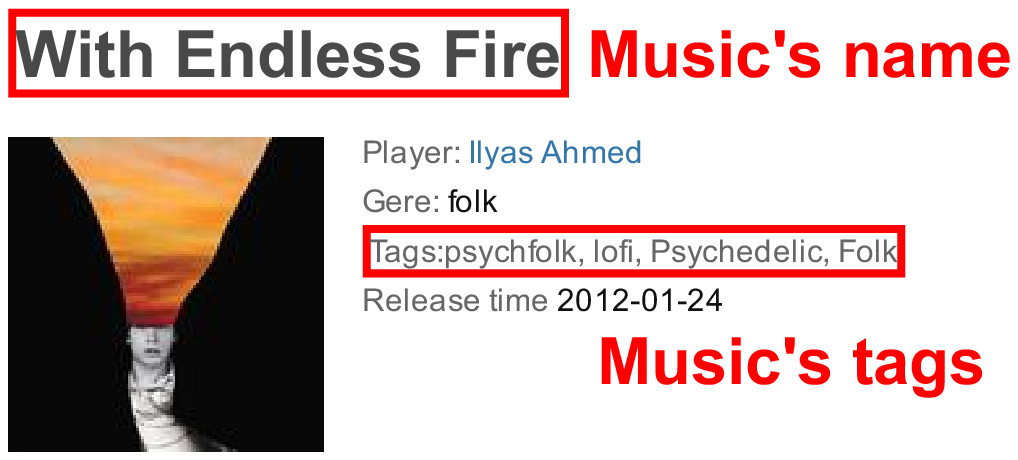}
}
\subfigure[An example of book's information.] { \label{figexample:3}
\includegraphics[width=0.47\columnwidth]{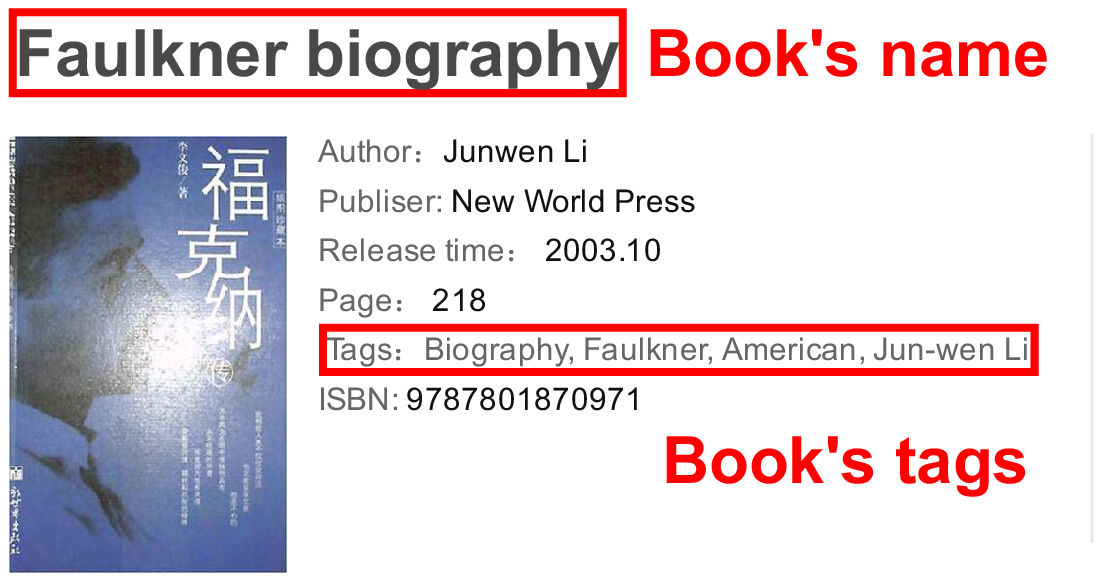}
} \vspace{-0.3cm}
\caption{Some Examples of Data in Douban. }
\label{figexample} \vspace{-0.3cm}
\end{figure}
\begin{table} [!th] \vspace{-0.3cm}
  \centering
  \caption{The statistics of seven pairs of data sets} \vspace{-0.3cm}
  \begin{tabular}{@{}ll@{}l@{}}
\hline
Source $\rightarrow$ Target &  \multicolumn{2}{l}{Statistics}\\
\hline
& \#user &$1,653$ \\
Movie\_category $\rightarrow$ Music\_tags& \#Movie\_category &$368,446$ \\
 Music\_tags $\rightarrow$ Movie\_category& \#Ave\_Movie\_category &$222.90$ \\
& \#Music\_tags &$9,229$ \\
& \#Ave\_Music\_tags &$5.58$ \\
\hline
& \#user &$1,653$ \\
Movie\_tags $\rightarrow$ Music\_tags& \#Movie\_tags &$317,742$ \\
Music\_tags $\rightarrow$ Movie\_tags& \#Ave\_Movie\_tags &$192.22$ \\
& \#Music\_tags &$9,229$ \\
& \#Ave\_Music\_tags &$5.58$ \\
\hline
& \#user &$1,653$ \\
Movie\_dir $\rightarrow$ Music\_tags& \#Movie\_dir &$373,164$ \\
Music\_tags $\rightarrow$ Movie\_dir& \#Ave\_Movie\_dir &$225.75$ \\
& \#Music\_tags &$9,229$ \\
& \#Ave\_Music\_tags &$5.58$ \\
\hline
& \#user &$423$ \\
Music\_tags $\rightarrow$ Book\_tags& \#Music\_tags &$2,474$ \\
Book\_tags $\rightarrow$ Music\_tags& \#Ave\_Music\_tags &$5.8$ \\
& \#Book\_tags &$25,342$ \\
& \#Ave\_Book\_tags &$59.91$ \\
\hline
& \#user &$3,750$ \\
Book\_tags $\rightarrow$ Movie\_category& \#Book\_tags &$155,974$ \\
Movie\_category $\rightarrow$ Book\_tags& \#Ave\_Book\_tags &$41.59$ \\
& \#Movie\_category &$471,810$ \\
& \#Ave\_Movie\_category &$125.82$ \\
\hline
& \#user &$3,750$ \\
Book\_tags $\rightarrow$ Movie\_tags& \#Book\_tags &$155,974$ \\
Movie\_tags $\rightarrow$ Book\_tags& \#Ave\_Book\_tags &$41.59$ \\
& \#Movie\_tags &$477,364$ \\
& \#Ave\_Movie\_tags &$127.29$ \\
\hline
& \#user &$3,750$ \\
Book\_tags $\rightarrow$ Movie\_dir& \#Book\_tags &$155,974$ \\
Movie\_dir $\rightarrow$ Book\_tags& \#Ave\_Book\_tags &$41.59$ \\
& \#Movie\_dir &$476,639$ \\
& \#Ave\_Movie\_dir &$127.10$ \\
\hline
\end{tabular} \label{tab:datastatistics}\\
$^*$A\_tags (category, dir) means A's tags (category, director and players). \vspace{-0.5cm}
\end{table}

\vspace{-0.3cm}
\subsection{Data Preparation}

We prepare the data sets by crawling the data from Douban, which is one of the most influential social-network service website in China, containing movie, music and book ratings of around millions of registered users. In Douban, users usually write a comment to movie, music or book after they have watched a movie, listened to a song or read a book. As shown in Fig.~\ref{figexample:a}, there is a user's watching list of movies, and each record contains the name of movie and the watching time (here we regard the time when user preforms the rating as the watching time). Also, there are descriptions of movie, music and book, whose examples are respectively shown in Figs.~\ref{figexample:1} to \ref{figexample:3}. For example, the description of a move contains movie's category, director and players, tags and so on.

We crawled the data from three domains of Movie (i.e., movie's category, director and players, and tags.), Music (i.e., music's tags.) and Book (i.e., book's tags.), and extracted the registered users who perform sequential behaviors on at least two domains. Finally, we constructed 14 transfer sequential recommendation problems (i.e., 7 pairs of data sets). For clarity, the statistics of seven pairs of data sets are summarized in Table~\ref{tab:datastatistics}. In this table, we provide the statistical information, including the number of users, the number of records (categories, director and players, tags), and the average number of records (categories, director and players, tags) for each user. From this table, we can find that movie data is much denser than book and music, and the music data is the most sparse.

\begin{table*}[th] \small
\newcommand{\tabincell}[2]{\begin{tabular}{@{}#1@{}}#2\end{tabular}}
  \centering
  \caption{Recommendation Performance on 7 Data Sets} \vspace{-0.3cm}
  \begin{tabular}{@{}c|cccccccc@{}}
\hline
\multicolumn{1}{c}{} & &\tabincell{c}{Music\_tags \\ $\rightarrow$Movie\_category } & \tabincell{c}{Music\_tags\\ $\rightarrow$ Movie\_tags } & \tabincell{c}{Music\_tags \\ $\rightarrow$ Movie\_dir} & \tabincell{c}{Music\_tags\\ $\rightarrow$ Book\_tags} & \tabincell{c}{Book\_tags \\$\rightarrow$ Movie\_category} & \tabincell{c}{Book\_tags\\ $\rightarrow$ Movie\_tags} &\tabincell{c}{Book\_tags \\ $\rightarrow$ Movie\_dir} \\
\hline
\multirow{7}{*}{MRR}&\tabincell{c}{OF}& $0.1406$ & $0.1234$ & $0.1204$ & $0.1915$ & $0.1945$ & $0.1770$ & $0.1730$ \\
& \tabincell{c}{OF\_U} &$0.1322$&$0.1140$ &$0.1072$&$0.1895$&$0.1815$&$0.1750$&$0.1663$\\
& \tabincell{c}{MC}& $\textbf{0.2733}$&$0.1397$&$0.1440$&$0.2052$&$0.2865$&$0.1845$&$0.1776$\\
&\tabincell{c}{MC\_U}& $0.2239$&$0.1322$&$0.1160$&$0.1884$&$0.2491$&$0.1693$&$0.1666$\\
&\tabincell{c}{NSM}& $0.1898$&$0.3217$&$0.3995$&$0.2253$&$0.2912$&$0.3578$&$0.4040$\\
&\tabincell{c}{NSM\_U} &$0.1830$&$0.3931$&$0.3981$&$0.2105$&$0.2521$&$0.3133$&$0.3618$\\
&\tabincell{c}{CDNST}&$0.2483$&$\textbf{0.4024}$&$\textbf{0.4746}$&$\textbf{0.2570}$&$\textbf{0.3414}$&$\textbf{0.3965}$&$\textbf{0.4162}$\\ \cline{2-9}
&\tabincell{c}{CDNST$^p$}&0.2570&0.4271& 0.5032&0.2675&0.3522& 0.4017& 0.4273\\
\hline
\multirow{7}{*}{nDCG@15} &\tabincell{c}{OF}& $0.1914$ & $0.1648$ & $0.1606$ & $0.2584$ & $0.2642$ &$0.2374$ & $0.2302$ \\
& \tabincell{c}{OF\_U} & $0.1798$ & $0.1540$ & $0.1467$ & $0.2398$ & $0.2505$ & $0.2397$ & $0.2227$ \\
& \tabincell{c}{MC}& $\textbf{0.3490}$ & $0.1848$ & $0.2870$ & $0.2817$ & $0.2906$ & $0.2464$ & $0.2347$ \\
&\tabincell{c}{MC\_U}&0.2957 & $0.1748$ & $0.1626$ &  $0.2504$ & $0.2751$ & $0.2285$ & $0.2225$ \\
&\tabincell{c}{NSM}&$0.2458$ & $0.3974$ & $0.4941$ & $0.2916$ & $0.3052$ & $0.4414$ & $0.4995$ \\
&\tabincell{c}{NSM\_U} &$0.2326$ & $0.4643$ & $0.4935$ & $0.2819$ & $0.2490$ & $0.3954$ & $0.4542$ \\
&\tabincell{c}{CDNST}&$0.2931$ &$\textbf{0.4778}$ &$\textbf{0.5500}$ &$\textbf{0.3529}$ &$\textbf{0.3823}$ &$\textbf{0.4829}$ &$\textbf{0.5148}$ \\
\cline{2-9}
&\tabincell{c}{CDNST$^p$}&0.3058 &   0.4930  &  0.5897  &  0.3614 &   0.3916  &  0.4875  &  0.5261\\
\hline
\multirow{7}{*}{p@3}&\tabincell{c}{OF}&$0.0866$ & $0.0768$ & $0.0706$ & $0.1154$ & $0.1356$ & $0.1231$ & $0.1226$ \\
& \tabincell{c}{OF\_U}& $0.0805$ & $0.0701$ & $0.0638$ & $0.0890$& $0.1210$ & $0.1173$ & $0.1139$ \\
& \tabincell{c}{MC}&$\textbf{0.2125}$ & $0.0935$ & $0.0749$ & $0.1244$ & $0.2248$ & $0.1303$ & $0.1256$ \\
&\tabincell{c}{MC\_U}& $0.1641$ & $0.0878$ & $0.0677$ & $0.1031$ & $0.1887$ & $0.1160$ & $0.1146$ \\
&\tabincell{c}{NSM}& $0.1435$ & $0.3040$ & $0.3897$ & $0.1553$ & $0.2581$ & $0.3396$ & $0.3851$ \\
&\tabincell{c}{NSM\_U} & $0.1420$ & $0.3781$ & $0.3903$ & $0.1202$ & $0.2365$ & $0.2971$ & $0.3510$ \\
&\tabincell{c}{CDNST}&$0.2028$ &$\textbf{0.3990}$ &$\textbf{ 0.4636}$ &$\textbf{0.2039}$ &$\textbf{ 0.3002}$ &$\textbf{0.3538}$ &$\textbf{0.4040}$ \\
\cline{2-9}
&\tabincell{c}{CDNST$^p$}&0.2166  &  0.4185 &   0.4792  & 0.2174  &  0.3137  &  0.3601 &   0.4109\\
\hline
\end{tabular}  \label{tab:res1} \vspace{-0.0cm}
\end{table*}

\begin{table*} [th] \small
\newcommand{\tabincell}[2]{\begin{tabular}{@{}#1@{}}#2\end{tabular}}
  \centering
  \caption{Recommendation Performance on The Other 7 Coupled Data Sets} \vspace{-0.3cm}
  \begin{tabular}{@{}c|cccccccc@{}}
    \hline
     \multicolumn{1}{c}{} & &\tabincell{c}{ Movie\_category \\ $\rightarrow$ Music\_tags  } & \tabincell{c}{Movie\_tags\\ $\rightarrow$ Music\_tags } & \tabincell{c}{ Movie\_dir \\ $\rightarrow$ Music\_tags} & \tabincell{c}{Book\_tags\\ $\rightarrow$ Music\_tags } & \tabincell{c}{Movie\_category\\$\rightarrow$ Book\_tags } & \tabincell{c}{Movie\_tags \\ $\rightarrow$ Book\_tags} &\tabincell{c}{ Movie\_dir \\ $\rightarrow$ Book\_tags} \\
     \hline
\multirow{7}{*}{MRR}&\tabincell{c}{OF}& $0.5093$ & $0.5093$ & $0.5093$ & $0.5149$ & $0.2483$ & $0.2483$ & $0.2483$ \\
& \tabincell{c}{OF\_U} & $0.3033$ & $0.3004$ & $0.3990$ & $0.1736$ & $0.1895$ & $0.1581$ & $0.1655$ \\
& \tabincell{c}{MC}& $0.5303$ & $0.5303$ & $0.5303$ & $0.5275$ & $0.2588$ & $0.2588$ & $0.2588$ \\
&\tabincell{c}{MC\_U}& $0.3921$ & $0.3174$ & $0.2969$ & $0.1869$ & $0.2414$ & $0.1853$ & $0.1822$ \\
&\tabincell{c}{NSM}&$\textbf{0.6842}$ &$\textbf{ 0.6842}$ &$\textbf{0.6842}$ &$\textbf{ 0.6891}$ &$\textbf{0.4031}$ &$\textbf{0.4031}$ &$\textbf{0.4031}$ \\
&\tabincell{c}{NSM\_U} & $0.1808$ & $0.3994$ & $0.3935$ & $0.3522$ & $0.2129$ & $0.3198$& $0.3676$ \\
&\tabincell{c}{CDNST}& $0.6745$ &$0.5755$ &$0.5659$ &$0.6628$ & $0.3616$ & $0.3347$ & $0.3350$ \\
\cline{2-9}
&\tabincell{c}{CDNST$^p$}&0.7196 &   0.7374   & 0.7063   & 0.7101 &   0.4459  &  0.5016  &  0.4282\\
\hline
\multirow{7}{*}{nDCG@15}&\tabincell{c}{OF}& $0.6145$ & $0.6145$ & $0.6145$ & $0.6584$ & $0.3323$ & $0.3323$ & $0.3323$ \\
& \tabincell{c}{OF\_U} & $0.4192$ & $0.4068$ & $0.3410$ & $0.4398$ &$0.2539$ & $0.2134$ & $0.2210$ \\
& \tabincell{c}{MC}& $0.6291$ & $0.6291$ & $0.6291$ & $0.6670$ & $0.3443$ & $0.3443$ & $0.3443$ \\
&\tabincell{c}{MC\_U} & $0.5019$ & $0.4234$ & $0.4055$ & $0.4504$ & $0.3173$ & $0.2531$ & $0.2491$ \\
&\tabincell{c}{NSM}&$\textbf{0.7599}$ &$\textbf{ 0.7599}$ &$\textbf{0.7599}$ &$\textbf{0.7657}$ &$\textbf{0.4989}$ &$\textbf{0.4989}$ &$\textbf{0.4989}$ \\
&\tabincell{c}{NSM\_U}& $0.2266$ & $0.4706$ & $0.4882$ &$0.4480$ & $0.2681$ & $0.4005$ & $0.4595$ \\
&\tabincell{c}{CDNST} &$0.7442$ & $0.6125$ &$0.5990$ &$0.7353$ &$0.4549$ &$0.4326$ &$0.4025$ \\
\cline{2-9}
&\tabincell{c}{CDNST$^p$}&0.7794 &   0.7930  &  0.7715  &  0.8061 &   0.5194 &   0.5147  &  0.5308\\
\hline
\multirow{7}{*}{p@3}&\tabincell{c}{OF}& $0.4600$ & $0.4600$ & $0.4600$ & $0.4319$ & $0.1830$ & $0.1830$ & $0.1830$ \\
&\tabincell{c}{ OF\_U} & $0.2219$ & $0.2281$ & $0.3555$ & $0.3131$ & $0.1319$ & $0.1058$ &$ 0.1137 $\\
& \tabincell{c}{MC}& $0.4832$ & $0.4832$ & $0.4832$ & $0.5266$ & $0.1928$ & $0.1928$ & $0.1928$ \\
&\tabincell{c}{MC\_U}& $0.3232$ & $0.2441$ & $0.2193$ & $0.3938$ & $0.1404$ & $0.1257$ & $0.1221$ \\
&\tabincell{c}{NSM}&$\textbf{0.6718}$ & $\textbf{0.6718}$ &$\textbf{ 0.6718}$ &$\textbf{ 0.6744}$ &$\textbf{0.3840}$ & $\textbf{0.3840}$ & $\textbf{0.3840}$ \\
&\tabincell{c}{NSM\_U} & $0.1486$ &$0.3843$ & $0.3831$ & $0.3349$ & $0.1855$ & $0.3024$ & $0.3556$ \\
&\tabincell{c}{CDNST}& $0.6655$ &$0.5495$ & $0.5425$ &$0.6350$ &$0.3529$ &$0.3228$ & $0.2898$ \\
\cline{2-9}
&\tabincell{c}{CDNST$^p$}&0.6852  &  0.7167  &  0.6950 &   0.7192  &  0.4108  &  0.4296 &  0.4007\\
\hline
\end{tabular}  \label{tab:res2} \vspace{-0.2cm}
\end{table*}

\vspace{-0.2cm}
\subsection{Evaluation Metrics and Baselines}
\subsubsection{Evaluation Metrics}
For all compared algorithms, they give a recommendation list of candidate choices with prediction probabilities, according to which we sort the candidate choices in descending order. In our experiments, the widely used evaluation metrics of nDCG~\cite{liu2009learning} , MRR~\cite{hanani2001information} and Precision~\cite{jarvelin2000ir} are adopted to evaluate the performance of all algorithms, and they are defined as follows,
\vspace{-0.3cm}
\begin{equation}\label{eq:ndcg}
\begin{aligned}
  \text{nDCG@k}=\frac{DCG_k}{IDCG_k},
 DCG_k=\sum_{i=1}^{k}\frac{2^{rel_i}-1}{\log_2{\left(i+1\right)}},
\end{aligned}
\end{equation}
\vspace{-0.3cm}
\begin{equation}\label{eq:mrr}
\hspace{-0.2cm}  \text{MRR}=\sum_{i=1}^{n}\frac{2^{rel_i}-1}{i}, \text{Precision@k}=\sum_{i=1}^{k}\frac{1}{i \cdot (2^{rel_i}-1)},
\end{equation} \vspace{-0.1cm}
where $i$ ranges over positions in the recommendation list and $rel_i$ reflects the preference of the $i$-th
items by the user.
Sort $rel_i$ by descending order and compute like Equation.~(\ref{eq:ndcg}), then we can obtain $IDCG$.
The result of dividing $DCG$ by $IDCG$ indicates the difference of recommendation order and true order.
When the actual result in prediction list is in the more front position, the value of MRR is larger; when the actual result in prediction list is in the more front position of $k$-position, nDCG@k and Precision@k are better.

\vspace{-0.2cm}
\subsubsection{Baselines}
We compare the proposed model CDNST with the following methods:
\begin{itemize}
  \item OF (Order by Frequency): OF method always gives a recommendation list according to the frequency in the individual's historical behavior sequence.
  \item OF\_U (Order by Frequency across domains): The only difference is that in OF\_U we compute the frequency in both source and target domains, while in OF only target domain is used.
  \item MC (Markon Chain)~\cite{markov1971extension}: The MC method models sequential behaviors in target domain by learning a transition graph and performing predictions.
  \item MC\_U (Markon Chain)~\cite{markov1971extension}: The MC method models sequential behaviors on both source and target domains by learning a transition graph and performing predictions.
  \item NSM (Novel Seeking Model)~\cite{ZhangYLX2014}: This is a data-driven model to predict the behaviour on target domain.
  \item NSM\_U:  We run the NSM model simply on both source and target domains, rather than in transfer manner.
\end{itemize}
We set the number of optional values for novelty-seeking level $K$ as 9 for both NSM and CDNST.

\vspace{-0.2cm}
\subsection{Experimental Results}

We first provide the experiments on all 14 transfer learning problems, and then show how the temporal property of data across different domain affects the proposed model CDNST.

\subsubsection{Effectiveness Results}

\begin{figure}[!th]
	\centering
       \includegraphics[width=7cm,height=5cm]{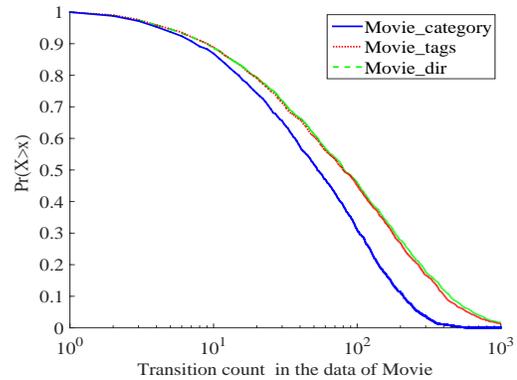}
   \vspace{-0.2cm}
	\caption{Long Tails of movie\_category, movie\_tags and movie\_dir.} \label{fig:longtails} \vspace{-0.4cm}
\end{figure}

These seven pairs of data sets are divided into two groups, i.e., the pair of $A \rightarrow B$ and $B \rightarrow A$ are put into different groups ($A$ and $B$ represent two domains), and all the results of the evaluation metrics of  nDCG@15, MRR and p@3 are shown in Table~\ref{tab:res1} and~\ref{tab:res2}. From these results, we have the following insightful observations, \\
$\bullet$  From Table~\ref{tab:res1}, we can find that our model CDNST outperforms all the baselines, except that on data set ``Music\_tags $\rightarrow$ Movie\_category", MC is slightly better than CDNST. And also, NSM performs at the second place. We try to investigate why MC performs the best on data set ``Music\_tags $\rightarrow$ Movie\_category". Fig.~\ref{fig:longtails} shows that the statistical information of Movie\_tags, Movie\_category and Movie\_dir. In Fig.~\ref{fig:longtails}, x-axis represents the number of transition status and y-axis represents the percentage of users whose transition status is larger than the given threshold value. From these results, we indeed find that the average number of transition status on Movie\_category is much smaller than the ones of Movie\_tags and Movie\_dir, which may be in favor of the MC method. \\
$\bullet$  Also, it is observed that incorporating the information from auxiliary source domain does not lead to the performance improvement, i.e., OF\_U, MC\_U and NSM\_U, which indicates that the previous models (i.e., OF, MC and NSM) can not effectively make full use of the auxiliary information. On the other hand, our model can benefit from the source domain to achieve significant improvement compared with NSM.\\
$\bullet$  Overall, NSM performs better than MC, and MC outperforms OF. \\
$\bullet$  From Table~\ref{tab:res2}, we find that the incorporating auxiliary information from source domain can not promote the performance of all algorithms on the second group of transfer learning problems, even the performances of baselines drop dramatically. After analyzing the data, we conjecture that the sequential property of auxiliary domain data affects the performance, which will be detailed in Section~\ref{sec:subanalysis}.

\vspace{-0.3cm}
\subsubsection{Analysis} \label{sec:subanalysis}

\begin{figure}[!th]
	\centering
		\includegraphics[width=9cm,height=3.5cm]{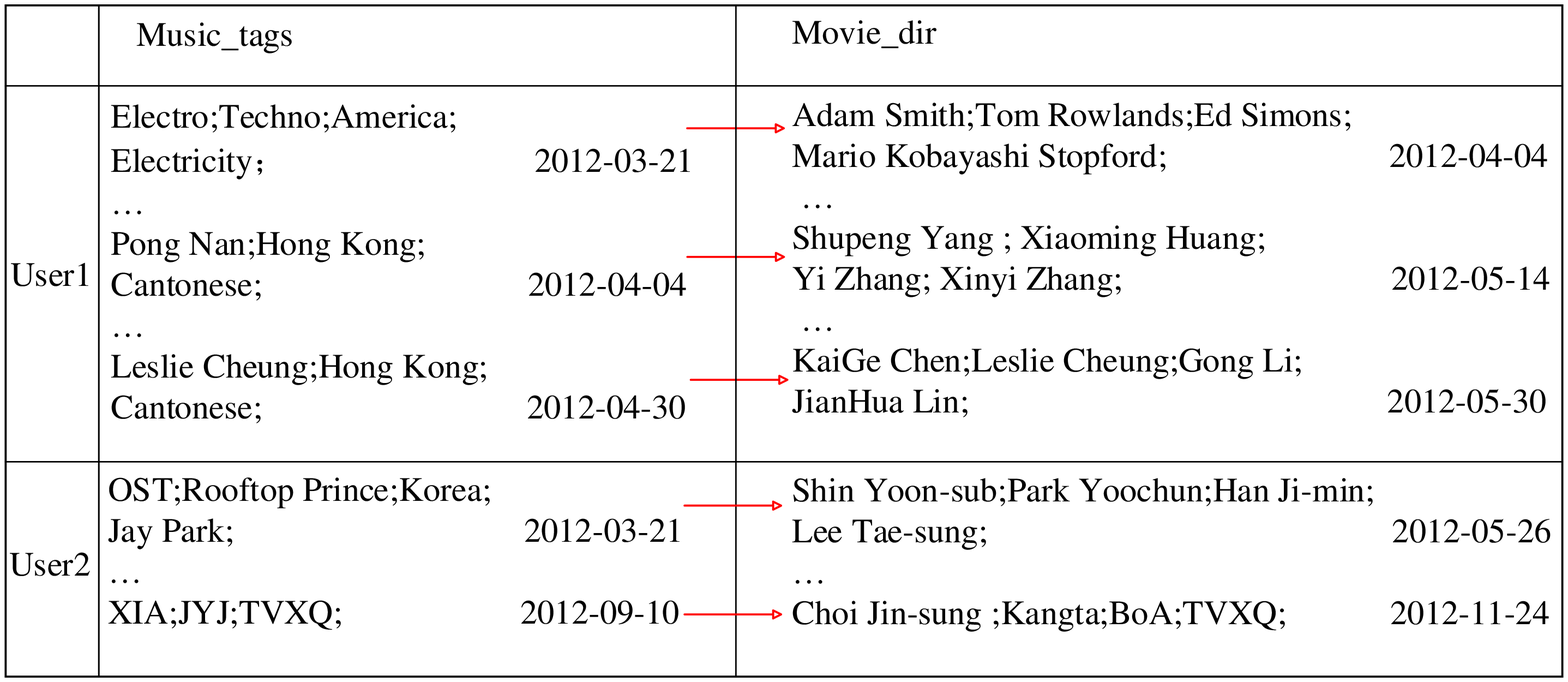} \vspace{-0.6cm} 
	\caption{Examples of Some Interesting Phenomenon of Users' Sequential Behaviors on Both Domains.}\label{fig:exampel}
\end{figure}

\begin{table*} [th] \small
\newcommand{\tabincell}[2]{\begin{tabular}{@{}#1@{}}#2\end{tabular}}
  \centering
  \caption{Recommendation Performance on The Other 7 Coupled Data Sets Advanced two months in source domain (The results of OF, MC and NSM are the same as the ones in Table~\ref{tab:res2}, which are omitted in this table.)} \vspace{-0.3cm}
  \begin{tabular}{@{}c|cccccccc@{}}
    \hline
     \multicolumn{1}{c}{} & &\tabincell{c}{ Movie\_category \\ $\rightarrow$ Music\_tags  } & \tabincell{c}{Movie\_tags\\ $\rightarrow$ Music\_tags } & \tabincell{c}{ Movie\_dir \\ $\rightarrow$ Music\_tags} & \tabincell{c}{Book\_tags\\ $\rightarrow$ Music\_tags } & \tabincell{c}{Movie\_category\\$\rightarrow$ Book\_tags } & \tabincell{c}{Movie\_tags \\ $\rightarrow$ Book\_tags} &\tabincell{c}{ Movie\_dir \\ $\rightarrow$ Book\_tags} \\
     \hline
\multirow{7}{*}{MRR}
& \tabincell{c}{OF\_U} &  $0.4056$&$0.4273$  & $0.3926$  & $0.4672$  & $0.1519$  &$0.1721$  & $0.1367$  \\
&\tabincell{c}{MC\_U}&$0.4295$   & $0.4412$  &$0.4093$   &$0.4620$   & $0.1982$  &$0.2073$  &$0.1842$   \\
&\tabincell{c}{NSM\_U} &  $0.6018$ & $0.6294$  & $0.5987$  &$0.6341$   & $0.3492$  &  $0.3263$ & $0.3392$  \\
&\tabincell{c}{CDNST}&$\textbf{0.7054}$ &$\textbf{ 0.7122 }$&$\textbf{0.6946}$&$\textbf{0.7067}$&$\textbf{0.4183}$  &$\textbf{0.4942}$ &$\textbf{0.4128}$  \\
\hline
\multirow{7}{*}{nDCG@15}
& \tabincell{c}{OF\_U} &$0.5561$   &$0.5726$   &  $0.5437$ & $0.5983$  &  $0.2572$ &  $0.2834$ &  $0.2163$\\
&\tabincell{c}{MC\_U}&$0.5836$ & $0.5982$  &$0.5727$  & $0.6068$  & $0.3064$  & $0.2985$  & $0.3183$     \\
&\tabincell{c}{NSM\_U}& $0.6283$ &  $0.6429$ & $0.6157$  & $0.6548$  &  $0.4851$ & $0.4746$  & $0.4954$ \\
&\tabincell{c}{CDNST} &$\textbf{0.7749}$  &$\textbf{ 0.7826}$  & $\textbf{0.7652}$ &$\textbf{0.7857}$  &$\textbf{ 0.5163}$ & $\textbf{0.5089}$ & $\textbf{0.5281}$ \\
\hline
\multirow{7}{*}{p@3}
&\tabincell{c}{ OF\_U} & $0.3843$  &$ 0.4058$  & $0.3651$  &$ 0.3039$ & $0.1288$  &$0.1325$   &$0.1158$   \\
&\tabincell{c}{MC\_U}& $0.4419$  &$0.4623$  &$0.4187$   & $0.3273$  & $0.1313$  &  $0.1458$ & $0.1207$  \\
&\tabincell{c}{NSM\_U} &  $0.6276$&  $0.6428$ & $0.6109$  &$0.3468$   & $0.3419$  & $0.3575$  &  $0.3249$ \\
&\tabincell{c}{CDNST}&$\textbf{0.6926}$   & $\textbf{0.7073}$ &$\textbf{ 0.6824}$  & $\textbf{ 0.7068}$ &$\textbf{0.4082}$  & $\textbf{0.4097}$ &$\textbf{0.3928}$   \\
\hline
\end{tabular}  \label{tab:res3} \vspace{-0.2cm}
\end{table*}

Overall, the results in Table~\ref{tab:res1} and~\ref{tab:res2} imply that the domains of Music and Book can help learn the model on Movie domain, and Music can help the learning of Book.
To intuitively show the temporal property of auxiliary domain data may affect the performances of all algorithms, we carefully investigate the characteristics of data set ``Music\_tags $\rightarrow$ Movie\_dir", and find some interesting phenomenon. Fig.~\ref{fig:exampel} lists some examples about the sequential behaviors of two users on both domains. For User 1, 1) he/she first listened to a song of the Chemical Brothers\footnote{https://en.wikipedia.org/wiki/The\_Chemical\_Brothers. Electro and Techno are members of the Chemical Brothers.} at time 2012/03/21 in the source domain, then later he/she would watch the movie about the Chemical Brother, e.g., ``The Chemical Brothers: Don't Think (2012)" at time 2012/04/04 in the target domain; 2) he/she first listened to the theme about the film of ``An Inaccurate Memoir" composed by Pong Nan at time 2012/04/04, then he/she would watched the movie of ``An Inaccurate Memoir" at time 2012/05/04; 3) he/she listened to the music of Leslie Cheung at time 2012/04/30, then later he/she would watch the movie ``Farewell My Concubine"\footnote{https://en.wikipedia.org/wiki/Farewell\_My\_Concubine\_(film)} with the player Leslie Cheung at time 2012/05/30. For User 2,  1) he/she first listened to the theme of Rooftop Prince at time 2012/03/21 in the source domain, then he/she would watch the movie of Rooftop Prince directed by Shin Yoon-sub at time 2012/05/26 in the target domain; 2) he/she listened to the music song by TVXQ at time 2012/09/10 in the source domain, then he/she would watch the movie of ``I AM.-SM Town Live World Tour in Madison Square Garden"\footnote{https://en.wikipedia.org/wiki/I\_AM.} played by the TVXQ in the target domain.

These examples may imply that given the source domain data Music\_tags, we can transfer the information to give better recommendation on target domain Movie\_dir. However in reverse, if we use Movie\_dir as source domain, which may not provide useful information for the recommendation on Music\_tags, since the related behaviors in Movie\_dir occur after the related ones in Music\_tags. Even worse, Movie\_dir may become noise data, which leads to the performance degrading. To further validate our analysis that the temporal property of source and target domain data affects the performance of the proposed model, we conduct simulation experiments on the second group of data sets. Specifically, we intentionally modify the occurring time of the behaviors in source domain, e.g., setting the occurring time by $\tau$ in advance ($\tau$ is set as two months in our experiments.), and conduct the experiments again on the second group of data sets. Table~\ref{tab:res3} records all the results, which show that the recommendation performance of all the algorithms becomes better, and our model CDNST again achieves the best results.

\begin{table*}[th]
\centering
\newcommand{\tabincell}[2]{\begin{tabular}{@{}#1@{}}#2\end{tabular}}
 \caption{\label{tab:test}The Relatedness on 7 Pairs of Data Sets} \vspace{-0.4cm}
 \begin{tabular}{ccccccc}
 \toprule
  \tabincell{c}{Music\_tags \\ $\rightarrow$Movie\_category } & \tabincell{c}{Music\_tags\\ $\rightarrow$ Movie\_tags } & \tabincell{c}{Music\_tags \\ $\rightarrow$ Movie\_dir} & \tabincell{c}{Music\_tags\\ $\rightarrow$ Book\_tags} & \tabincell{c}{Book\_tags \\$\rightarrow$ Movie\_category} & \tabincell{c}{Book\_tags\\ $\rightarrow$ Movie\_tags} &\tabincell{c}{Book\_tags \\ $\rightarrow$ Movie\_dir} \\
  \midrule
 $0.3955$&	$0.4213$ &	$0.2576$&	$0.3818$& $0.3366$&	$0.2008$&	$0.2412$\\
 \hline
 \tabincell{c}{ Movie\_category\\ $\rightarrow$Music\_tags } & \tabincell{c}{Movie\_tags \\ $\rightarrow$ Music\_tags} & \tabincell{c}{Movie\_dir \\ $\rightarrow$ Music\_tags } &
 \tabincell{c}{Book\_tags\\  $\rightarrow$ Music\_tags } &
  \tabincell{c}{ Movie\_category \\ $\rightarrow$ Book\_tags} & \tabincell{c}{Movie\_tags\\ $\rightarrow$ Book\_tags }
  &\tabincell{c}{ Movie\_dir\\ $\rightarrow$ Book\_tags} \\
 \midrule
 $0.3416$	&$0.3261$&	$0.1551$	&$0.3128$&	$0.2910$	&$0.1493$&	$0.1932$\\
  \midrule
  $0.4002 \uparrow$&	$0.3778 \uparrow$&	$0.2289 \uparrow$ &	$0.4374 \uparrow$ &	$0.3541 \uparrow$&	$0.1969 \downarrow$	& $0.2540 \uparrow$\\
  \bottomrule
 \end{tabular}\vspace{-0.2cm}
\end{table*}

Obviously, the transfer learning problem for sequential data is different from previous works, since it is directed. As we know, almost all the previous transfer learning algorithms are undirected, which are assumed to work well on both cases $A \rightarrow B$ and $B \rightarrow A$. This may lead to the failure when the problem does not satisfy the temporal property. Is it possible to propose a relatedness measure to judge whether a transfer learning problem is suitable to our model CDNST? To this end, we propose an effective measure $Sim$ by incorporating the external web data, and $Sim(A \rightarrow B) \neq Sim(B \rightarrow A)$. We hope when $Sim(A \rightarrow B) > Sim(B \rightarrow A) $, our model CDNST can make sense, and vice versa.

\begin{figure}[!th] \vspace{-0.5cm}
	\begin{center}
       \includegraphics[scale=0.4]{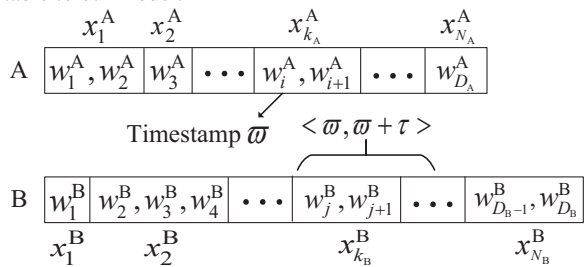}
	\end{center}
	\vspace{-0.5cm}
	\caption{The chronological order of actions from two domains for one user.}\label{fig:graphicalmodel2} \vspace{-0.4cm}
\end{figure}

There are keywords in the context information, so before formally defining the measure $Sim$, we will first introduce how to compute the similarity of two keywords and the relatedness of behaviours from different domains for each user. For each keyword of both domains, we use it as a query and crawl the top 100 results from the search engine (e.g., Baidu and Google.) to form a corpus. Then, we can convert a keyword to a vector using the word2vec technique\footnote{http://spark.apache.org/docs/1.3.1/mllib-feature-extraction.html\#word2vec.}, and $Sim(w_i,w_j) = \frac{v_i^{\top} v_j}{\sqrt{v_i^{\top} v_i} \sqrt{v_j^{\top} v_j}}$, where $w$ denotes a keyword, $v$ denotes its corresponding vector, and $v^{\top}$ denotes the transposition of $v$ (the number of dimension is set as 50 in the experiments.). As shown in Fig.~\ref{fig:graphicalmodel2}, we sort the keywords of user $u$ in $A$ domain $(w_1^A,w_2^A,\cdots,w_{D_A}^A)$ and in $B$ domain $(w_1^B,w_2^B,\cdots,w_{D_B}^B)$ with chronological order, where $D_A$ and $D_B$ respectively denote the number of keywords in domains $A$ and $B$, and then the relatedness of behaviours in the case of $A \rightarrow B$ for each user is defined as,
\begin{eqnarray} \label{eq:userRelatenessmeasure}
\begin{split}
Sim(A_u \rightarrow B_u) = \frac{1}{N_p} \sum_{i=1}^{D_A}\sum_{w_j^{B}\in \mathcal{W}_i^B} Sim(w_i^A,w_j^B),
\end{split}
\end{eqnarray}
where if the timestamp of $w_i^A$ is $\varpi$, $\mathcal{W}_i^B$ denotes the set of keywords in $B$ domain, whose timestamp is between $\varpi$ and $\varpi + \tau$, and $N_p$ is the total number of keyword pairs for user $u$. Finally, we are ready to define $Sim(A \rightarrow B)$,
\begin{eqnarray} \label{eq:domainRelatenessmeasure}
\begin{split}
Sim(A \rightarrow B) =\frac{1}{N}\sum_u Sim(A_u \rightarrow B_u).
\end{split}
\end{eqnarray}
where $N$ is the number of users.
We compute the relatedness measure of all 14 problems and 7 new constructed problems in Section~\ref{sec:subanalysis}, and all results are recorded in Table~\ref{tab:test}. From these results, we can find that the values of relatedness measure on the first group (in the second row) are all larger than the ones on the second group (in the fourth row), which is coincident with our analysis. Also on the new constructed problems, the values of relatedness measure are significantly increased. Therefore, we can adopt this relatedness measure $Sim$ to judge whether a transfer learning problem is suitable to our model.

\vspace{-0.2cm}
\subsubsection{Transfer for Personalized Recommendation} \label{sec:subpersonalized}
Furthermore, we can adopt the proposed relatedness measure $Sim$ to make effective transfer on the user level recommendation, which is very useful for personalized recommendation. Specifically, according to Equation.~(\ref{eq:userRelatenessmeasure}) the users are chosen who are suitable for the transfer scenario $A \rightarrow B$ (i.e., $Sim(A_u \rightarrow B_u) > Sim(B_u \rightarrow A_u)$) or $B \rightarrow A$ (i.e., $Sim(B_u \rightarrow A_u) > Sim(A_u \rightarrow B_u)$), then we can run CDNST on these corresponding users on seven pairs of data sets. The results are shown in the last row of each metric in Table~\ref{tab:res1} and~\ref{tab:res2} (Our model is denoted as CDNST$^p$ for this personalized recommendation). From these results, we can find that CDNST can obtain additional improvement compared with the one transfer on domain level, which again indicate the effectiveness of the proposed relatedness measure.

\vspace{-0.3cm}
\section{Conclusions and Remarks} \label{sec:conclusions}

In this paper, we propose a new cross-domain recommendation algorithm, in which the novelty-seeking trait of users are shared across source and target domains for effective knowledge transfer. To validate the effectiveness of the proposed model, we first crawl three domains of data sets from the well-known Chinese social-media platform Douban, and construct 14 transfer recommendation problems. The experiments show that our model is more accurate, when the source and target domain data satisfy the sequential property, i.e., the related behaviors in source domain occur before the related ones in target domains. This may be a new cross-domain recommendation problem, which we call it
sequential recommendation. In the future, we will aim to propose new transfer recommendation model to address this problem.

\vspace{-0.2cm}
\bibliographystyle{ACM-Reference-Format}
\bibliography{sigproc}  

\balance


\end{document}